\documentclass[twocolumn,showpacs,nofootinbib,preprintnumbers,prd]{revtex4-1}

\usepackage{amsmath}
\usepackage{amsfonts}
\usepackage{amssymb}
\usepackage{graphicx}
\usepackage{changepage}
\usepackage[titletoc]{appendix}
\usepackage{color}
\usepackage{hyperref}
\usepackage{cleveref}
\usepackage[rightcaption]{sidecap}
\usepackage{subfigure}
\usepackage{comment}

\usepackage{mathtools}

\usepackage{dcolumn}

\usepackage{array}
\usepackage{ctable}
\usepackage{multirow}
\usepackage{siunitx}
\usepackage{longtable}
\usepackage{tabularx}
\usepackage{booktabs}
\usepackage{supertabular}
\usepackage{verbatim}

\graphicspath{{Graphics/}}

\def\be{\begin{equation}}
\def\ee{\end{equation}}
\def\bea{\begin{eqnarray}}
\def\eea{\end{eqnarray}}

\definecolor{vividviolet}{rgb}{0.62, 0.0, 1.0}
\definecolor{amaranth}{rgb}{0.9, 0.17, 0.31}
\definecolor{palatinateblue}{rgb}{0.15, 0.23, 0.89}
\definecolor{brightpink}{rgb}{1.0, 0.0, 0.5}
\definecolor{cornflowerblue}{rgb}{0.39, 0.58, 0.93}
\definecolor{deepcarminepink}{rgb}{0.94, 0.19, 0.22}
\definecolor{radicalred}{rgb}{1.0, 0.21, 0.37}

\hypersetup{ linktoc=all,
    colorlinks, linkcolor={palatinateblue},
    citecolor={brightpink}, urlcolor={amaranth}
}

\begin{document}

\title{Cosmological consequences of scale-dependent Barrow-Tsallis entropy}

\author{Giuseppe Gaetano Luciano}
\email{giuseppegaetano.luciano@udl.cat}
\affiliation{Departamento de Qu\'{\i}mica, F\'{\i}sica y Ciencias Ambientales y del Suelo, Escuela Polit\'ecnica Superior - Lleida, Universidad de Lleida, Av. Jaume II, 69, 25001 Lleida, Spain.}

\author{Orlando Luongo}
\email{orlando.luongo@unicam.it}
\affiliation{Universit\`a di Camerino, Divisione di Fisica, Via Madonna delle carceri 9, 62032 Camerino, Italy.}
\affiliation{Department of Nanoscale Science and Engineering, University at Albany-SUNY, Albany, New York 12222, USA.}
\affiliation{INFN, Sezione di Perugia, Perugia, 06123, Italy.}
\affiliation{INAF - Osservatorio Astronomico di Brera, Milano, Italy}
\affiliation{Al-Farabi Kazakh National University, Al-Farabi av. 71, 050040 Almaty, Kazakhstan.}

\author{Marco Muccino}
\email{marco.muccino@lnf.infn.it}
\affiliation{Universit\`a di Camerino, Divisione di Fisica, Via Madonna delle carceri 9, 62032 Camerino, Italy.}
\affiliation{Al-Farabi Kazakh National University, Al-Farabi av. 71, 050040 Almaty, Kazakhstan}
\affiliation{INAF - Catania Astrophysical Observatory, Via S.Sofia 79, 95123 Catania, Italy.}
\affiliation{ICRANet, Piazza della Repubblica 10, 65122 Pescara, Italy.}

\date{\today}
 
\begin{abstract}

We investigate an extended cosmological scenario based on the Barrow-Tsallis entropy, incorporating a varying (i.e., energy-scale-dependent) anomalous dimension. This behavior is reminiscent of quantum gravity and effective field theory settings, where the relevant couplings acquire a nontrivial scale dependence. By applying the gravity-thermodynamic conjecture on the apparent horizon of a flat Friedmann-Robertson-Walker Universe, we derive the corresponding modified cosmological equations. The standard dynamics is recovered as a limiting case when the Barrow-Tsallis entropy reduces to the conventional Bekenstein-Hawking form.
The proposed model is tested against a combination of early- and late-time datasets, including observational Hubble parameter measurements, the Pantheon+ catalog of Type~Ia supernovae, the second data release of DESI baryon acoustic oscillations and cosmic microwave background constraints. 
Using the Bayesian Information Criterion, we finally compare the fitting performance of our framework with that of the $\Lambda$CDM paradigm. While the latter remains mildly favored, our model is shown to be fully compatible
with current observations within suitable regions of the parameter space,
unveiling a richer phenomenology that points towards a possible alleviation of the Hubble tension.
\end{abstract}

\maketitle
\tableofcontents 

\section{Introduction}
\label{Intro}

Formulating a unification scheme of quantum field theory and general relativity remains an open challenge of modern theoretical physics \cite{Kiefer:1993fg,Donoghue:1994dn}. Indeed, despite significant progress, a complete theory of quantum gravity is still lacking \cite{2005NJPh....7..198A,Ashtekar:2014kba} and existing proposals are typically affected by conceptual limitations and/or theoretical issues~\cite{Rovelli:2000aw,Addazi:2021xuf}.

Remarkably, the renewed interest in black hole (BH) physics has shed light on the highly energetic regimes where classical gravity is expected to break down \cite{Hawking:1975vcx,Bekenstein:1973ur}. In particular, black hole thermodynamics suggests that gravitational dynamics may admit an effective thermodynamic description, offering an alternative perspective on the microscopic structure of spacetime \cite{Jacobson:1995ab,Padmanabhan:2009vy}. Motivated by this connection, considerable attention has been devoted to extending the standard Boltzmann-Gibbs framework\footnote{Such generalizations are particularly suited to systems with long-range interactions, strong correlations, or fractal structures, where the standard assumption of extensivity may no longer hold.}, leading to several non-extensive entropy formalisms, including the Tsallis~\cite{Tsallis:1987eu,Tsallis:2009,Tsallis:2013}, R\'enyi~\cite{renyi1961entropy,Jizba:2002um}, Sharma-Mittal~\cite{Sharma1975} and Kaniadakis~\cite{kaniadakis2001non,Kaniadakis:2002zz,Luciano:2022eio} formulations.

Within this broader framework, a particularly relevant example is provided by the Barrow entropy~\cite{Barrow:2020tzx}. 
In this approach, quantum fluctuations induce a fractal-like deformation of the BH horizon geometry \cite{Jalalzadeh:2021gtq,Jalalzadeh:2022uhl}, leading to a power-law correction to the Bekenstein-Hawking entropy~\cite{Barrow:2020tzx} through the introduction of a constant \emph{anomalous dimension}, denoted by $\Delta$. Interestingly,  in the perturbative regime 
the anomalous dimension 
satisfies $|\Delta| \ll 1$, corresponding to small deviations from the area law. In this limit, the Barrow entropy reduces to the standard Bekenstein-Hawking expression supplemented by a logarithmic correction, as predicted in several quantum gravity scenarios~\cite{Kaul:2000kf,Carlip:2000nv,Adler:2001vs,Banerjee:2011jp}.
Furthermore, it is worth noting that, although the Barrow entropy shares the same functional form as the Tsallis $\delta$-entropy~\cite{Tsallis:1987eu,Tsallis:2009,Tsallis:2013}, the two formulations have distinct physical origins, with the latter emerging from the framework of nonextensive statistical mechanics.

More recently, the application of this thermodynamic recipe to cosmology has attracted growing attention~\cite{Jacobson:1995ab} through the so-called \emph{gravity-thermodynamics conjecture}~\cite{Padmanabhan:2003gd}, which establishes a deep connection between gravitational dynamics and thermodynamic laws, particularly in the context of horizon thermodynamics~\cite{Hayward:1997jp,Padmanabhan:2009vy,Cai:2006rs,Gibbons:1977mu}. Specifically, it has been shown that the Friedmann equations can be derived from the first law of thermodynamics, combined with a Clausius-like heat-entropy relation, by modeling the Universe as a thermodynamic system bounded by the apparent horizon and adopting the Bekenstein-Hawking area law as the underlying entropy functional~\cite{Frolov:2002va,Cai:2005ra,Akbar:2006kj}. 
In this thermodynamic picture, any modification of the entropy-area relation naturally leads to a generalized class of Friedmann equations~\cite{Tavayef:2018xwx,Saridakis:2020lrg,Lymperis:2018iuz,Lymperis:2021qty,Luciano:2022ely,Barboza:2014yfe,Nunes:2015xsa,Barrow:2020tzx,Saridakis:2020zol,Nojiri:2021jxf,Saridakis:2020lrg,Adhikary:2021xym,Sheykhi:2021fwh,DiGennaro:2022ykp,Dabrowski:2020atl,Mamon:2020spa,Jusufi:2021fek,Luciano:2022pzg,Luciano:2022hhy,Luciano:2023wtx,Jizba:2024klq,Luciano:2023roh}, with the potential to alleviate current cosmological tensions~\cite{Basilakos:2023kvk,Yarahmadi:2024oqv}.

In the specific case of Barrow entropy,  the modified Friedmann equations contain additional contributions that effectively behave as a dark-energy sector, while the standard $\Lambda$CDM cosmology is consistently recovered in the limit $\Delta=0$~\cite{Saridakis:2020zol}.
Attempts to compare the anomalous dimension with cosmological probes have been predominantly explored under the assumption that $\Delta$ is constant, with only a few notable exceptions~\cite{DiGennaro:2022ykp,Nojiri:2019skr,Basilakos:2023seo}. These studies, however, either explore running entropy exponents in the context of generalized nonextensive thermodynamics or consider holographic dark-energy realizations, leaving the observational viability of scale-dependent Barrow cosmology largely unexplored.

Motivated by the above discussion, in this work we investigate a cosmological scenario based on a Barrow-like deformation of the Bekenstein-Hawking entropy with a dynamical anomalous dimension. Indeed, if $\Delta$ is interpreted as encoding quantum imprints, it appears physically more natural to promote it to a scale-dependent quantity, allowing it to evolve across different cosmic epochs. Such a behavior naturally captures the interplay between ultraviolet quantum-gravitational effects, expected to dominate at early times, and infrared contributions, relevant at late times. Accordingly, we introduce a redshift-dependent Barrow exponent whose evolution is inspired by quantum-gravity considerations. More specifically, its functional form is motivated by the logarithmic corrections that, as discussed above, naturally emerge in quantum-gravity frameworks. We further assume that the evolving anomalous dimension is governed by the Hubble parameter, which provides a natural measure of the cosmological scale. This choice translates the underlying logarithmic correction into a scale-dependent evolution, leading to a stronger quantum-gravitational imprint at earlier epochs.

We then derive the corresponding modifications to the Friedmann equations, characterizing an exact solution of the continuity equation.
Accordingly, we show that this framework gives rise to an evolving dark-energy component at late times, which remains subdominant at high redshifts. To assess the viability of our scenario, we compare its predictions against a combination of datasets, including observational Hubble data (OHD), the Pantheon+ sample of type Ia supernovae (SNe~Ia), the second data release (DR2) of DESI baryon acoustic oscillations (BAO), and cosmic microwave background (CMB) constraints. As a byproduct of our analysis, we infer the corresponding constraints on the free parameters of the model and explore the phenomenological implications of this extended scenario. In particular, 
we focus on its potential characteristics
to improve the standard cosmological framework, predicting
a viable dark energy evolution and enlarging the
expected Hubble constant at late-times. We also discuss possible physical refinements of the present framework and outline natural extensions of the model to more general scenarios.

The paper is structured as follows. In Sec.~\ref{ModCosm}, we employ the gravity-thermodynamics conjecture to derive the modified Friedmann equations. In Sec.~\ref{sezione3}, we first review the Barrow entropy and then introduce our evolving paradigm, proposing a scale-dependent anomalous dimension. In Sec.~\ref{NumAn}, we describe the datasets used in the numerical analysis and derive the corresponding cosmological observables relevant for parameter estimation. Our physical results are discussed in Sec.~\ref{sezione5}, where we critically analyze and interpret the resulting constraints. In particular, Sec.~\ref{HubTen} is devoted to the Hubble tension and its possible alleviation within our framework. Finally, conclusions and perspectives are summarized in Sec.~\ref{Concl}.
We work in natural units with $\hbar=c=k_{\rm B}=1$

\section{Gravity-thermodynamic conjecture and modified cosmology}
\label{ModCosm}

To formulate the gravity-thermodynamics conjecture, we follow Ref.~\cite{Cai:2005ra}, according to which the starting point is the application of the first law of thermodynamics to the apparent horizon of the Friedmann-Robertson-Walker (FRW) Universe.
This horizon is defined as a marginally trapped surface with vanishing expansion, whereas the first law reads
\begin{equation}
\label{ftl}
    dE= T dS + W dV\,,
\end{equation}
where $E$ denotes the total energy contained within the apparent horizon, which is characterized by temperature $T$ and entropy $S$, and $dE$ represents the infinitesimal change in energy over the time interval $dt$. The quantity $W$ is the work density associated with a change in the horizon radius, and $V$ is the volume enclosed by the apparent horizon.

An expanding homogeneous and isotropic universe in $(3+1)$ dimensions is described by the FRW metric
\be
\label{FRW}
ds^2\,=\,h_{ab}\hspace{0.2mm}dx^{a}dx^{b}+\tilde r^2\left(d\theta^2+\sin^2\theta\,d\phi^2\right),
\ee
where $\tilde{r} = a(t)r$ represents the proper distance at cosmic time $x^0=t$, with $x^1=r$ denoting the comoving radial coordinate, while $h_{ab}=\mathrm{diag}[-1,a^2/(1-kr^2)]$ is the metric of the $(1+1)$-dimensional subspace. The curvature parameter $k=\{0,+1,-1\}$ corresponds to flat, closed, and open spatial geometries, respectively.

In cosmological contexts, applications of the gravity-thermodynamics conjecture are carried out by identifying the apparent horizon as the relevant holographic boundary. Hence, since the radius of this horizon is determined by the condition
\begin{equation}
  h^{ab} \partial_a \tilde{r} \partial_b \tilde{r} = 0,  
\end{equation}
one obtains~\cite{Cai:2005ra}
\begin{equation}
    \label{rA}
    \tilde r_A =\frac{1}{\sqrt{H^2+k/a^2}}\,,
\end{equation}
where $H=\dot a/a$ is the Hubble parameter, with the dot denoting the derivative with respect to cosmic time.
Under the assumption of a spatially flat Universe, as supported by recent CMB and BAO observations~\cite{Planck:2018vyg}, Eq.~\eqref{rA} simplifies to the Hubble radius, $\tilde{r}_A = 1/H$.

Assuming that the energy-momentum distribution of the Universe is described by the perfect-fluid energy-momentum tensor, $T_{\mu\nu} = (\rho + p)\,u_\mu u_\nu + p g_{\mu\nu}$,
where $u_\mu$ is the four-velocity of the fluid, while $\rho$ and $p$ denote its energy density and pressure, respectively, the conservation law $\nabla_\mu T^{\mu\nu}=0$ leads to the continuity equation
\begin{equation}
\label{cont}
     \dot{\rho} + 3H(\rho + p) = 0\,.
\end{equation}
Furthermore, following Refs.~\cite{Bak:1999hd,Hayward:1997jp,Awad:2014bta}, the work density evaluated at the apparent horizon is defined as
\begin{equation}
W = -\frac{1}{2} T^{ab} h_{ab} = \frac{1}{2}(\rho-p)\,.
\end{equation}

At this stage, we need to assign both entropy and temperature to the holographic boundary\footnote{These quantities are typically derived from BH thermodynamics, where the role of the cosmological apparent horizon is replaced by the BH event horizon.}.
In the BH context, it is well established that the temperature is determined by the surface gravity at the horizon, namely
\begin{equation}
\label{Tgen}
    T=\frac{\kappa}{2\pi}=
    \frac{1}{2\pi}\frac{1}{2\sqrt{-h}}\partial_a\left(\sqrt{-h}h^{ab}\partial_b\tilde r\right).
\end{equation}
Substituting the FRW metric~\eqref{FRW} into Eq.~\eqref{Tgen}, one obtains the temperature associated with the apparent horizon~\cite{Cai:2005ra,Awad:2014bta},
\begin{equation}
\label{T}
    T=-\frac{1}{2\pi\tilde r_A} \left(1-\frac{\dot{\tilde r}_A}{2H\tilde r_A}\right).
\end{equation}
Moreover, the entropy associated with the BH horizon is given by the Bekenstein-Hawking area law,
\begin{equation}
S = \frac{A}{4G},
\end{equation}
where the horizon area is
\begin{equation}
  A = 4\pi \tilde{r}_A^2,
\end{equation}
while the energy enclosed within the apparent horizon and the corresponding volume are given by
\begin{equation}
 E=\rho V,\qquad V=\frac{4}{3}\pi \tilde r_A^3\,.
\end{equation}
Thus, we end up with 
\begin{eqnarray}
\label{dE}
dE&=& 4\pi \tilde  r_A^2 \rho d\tilde r_A-4\pi H \tilde 
r_A^3\left(\rho+p\right)dt\,,
\\[2mm]
T dS &=& -\frac{1}{G} \left(1-\frac{\dot 
{\tilde r}_A}{2H\tilde r_A}\right) d\tilde r_A\,,
\\[2mm]
\label{dS}
W dV &=& 2\pi \tilde r_A^2\left(\rho-p\right)d\tilde r_A\,,
\label{dV}
\end{eqnarray}
where, in Eq.~\eqref{dE}, we used Eq.~\eqref{cont}.

By substituting Eqs.~\eqref{dE}--\eqref{dV} into Eq.~\eqref{ftl}, one straightforwardly obtains
\begin{equation}
    \frac{d\tilde{r}_A}{\tilde{r}_A^3} = 4\pi G\left(\rho + p\right)H\, dt,
\end{equation}
which can be further manipulated, using $\tilde r_A=1/H$, to recover the second Friedmann equation,
\begin{equation}
\label{F1}
    \dot H = -4\pi G\left(\rho+p\right).
\end{equation}
By employing Eq.~\eqref{cont}, the above relation can then be integrated to yield the first Friedmann equation,
\begin{equation}
\label{F2}
    H^2 = \frac{8\pi G}{3} \rho + \frac{\Lambda}{3}\,,
\end{equation}
where the integration constant $\Lambda$ can be interpreted as the cosmological constant~\cite{Padmanabhan:2010zzb}.

\section{Generalizing the Barrow entropy with a running term}\label{sezione3}

The Barrow entropy provides a simple extension of the Bekenstein-Hawking area law by introducing a deformation parameter that encodes possible quantum-gravitational modifications of the horizon geometry. Its standard expression can be written as~\cite{Barrow:2020tzx}
\begin{equation}
S_B = \left( \frac{A}{A_0} \right)^{1 + \Delta/2}\,,
\label{Barrow}
\end{equation}
where $A$ denotes the horizon area, which can generally be expressed as $A=4\pi L^2$ in terms of the associated horizon length scale $L$, while $A_0\equiv4G$ is the Planck area that sets the entropy normalization. The parameter $\Delta$, referred to as the \emph{anomalous dimension}, is usually assumed to be constant and takes values in the interval $0\leq\Delta\leq1$. The limiting cases $\Delta=0$ and $\Delta=1$ correspond to the standard Bekenstein-Hawking entropy of a smooth horizon and to a horizon with maximal fractal deformation, respectively.

Although Barrow originally introduced Eq.~\eqref{Barrow} in the context of a simple sphereflake-like BH model, thereby assuming a positive value of $\Delta$, subsequent theoretical and phenomenological developments have suggested that this restriction may be unnecessarily limiting. For instance, surface geometries exhibiting internal voids or porous structures can be characterized by negative fractal dimensions~\cite{tang2012fractal,xu2008developing}. Additional support for extending $\Delta$ to negative values arises from quantum field theory and renormalization-group arguments~\cite{Dagotto:1989gp}. Accordingly, the admissible range of the anomalous dimension can be generalized to~\cite{Jizba:2024klq,Anagnostopoulos:2020ctz,Luciano:2025hjn}
\begin{equation}
\Delta \in (-1,1].
\end{equation}
The above extension also motivates a more general perspective, in which the anomalous dimension is promoted from a constant to a scale-dependent quantity. In such a case, $\Delta$ is naturally expected to evolve throughout the expansion history of the Universe, with direct consequences for the form of the modified Friedmann equations.

Having established the above framework, we are now in a position to extend the previous derivation by incorporating the Barrow entropy with a scale-dependent anomalous dimension. Following the approach of Ref.~\cite{Awad:2014bta}, we introduce the generalized entropy
\begin{equation}
\label{genent}
    S = \frac{f(A)}{A_0}\,,
\end{equation}
where $f(A)$ is a generalized function of the apparent-horizon area that reproduces the standard Barrow entropy in Eq.~\eqref{Barrow} through
\begin{equation}
\label{Bent}
f(A)=\dfrac{A^{1+\Delta/2}}{A_0^{\Delta/2}}\,.
\end{equation}

Using the first law of thermodynamics and following the same procedure as outlined above, one obtains
\begin{equation}
    \frac{df}{dA} \frac{d\tilde{r}_A}{\tilde{r}_A^3} = 4\pi G\left(\rho + p\right)H\, dt\,.
\end{equation} 
Further substitution of Eq.~\eqref{rA} yields 
\begin{equation}
\label{ModF1}
    \dot H\, \frac{df}{dA}  =-4\pi G\left(\rho+p\right).
\end{equation}
As the next step, substituting the continuity equation, Eq.~\eqref{cont}, and integrating, we obtain the generalized first Friedmann equation in the following form
\begin{equation}
    -4\pi\int^A\frac{df}{dA}\hspace{0.3mm}\frac{dA}{A^2}=\frac{8\pi G}{3} \rho\,,
    \label{ModF2}
\end{equation}
in which we adopted  $A=4\pi/H^2$. 

Let us now specialize Eq.~\eqref{ModF2} to the Barrow model of Eq.~\eqref{Bent} and introduce a new paradigm in which the anomalous dimension $\Delta$ is allowed to vary with the scale. We motivate this extension on the basis of three main arguments, which are discussed below:

\begin{enumerate}
    \item Within the cosmological context, the energy scale can be  quantified by employing the Hubble parameter $H(z)$ as the Hubble radius, i.e., a prototype of cosmic distance reduces to the apparent horizon as $k=0$.  Hence, in order to account for a varying entropic exponent, it is convenient to introduce the dimensionless quantity 
    \begin{equation}
    x \equiv [H_1/H(z)]^2,     
    \end{equation}
    with  $H_1=\sqrt{\pi/G}$ setting the reference scale \cite{Nojiri:2019skr}. 
Hence, Eq.~\eqref{ModF2} can be equivalently recast into
\begin{equation}
-\!\int\! x^{\frac{\Delta(x)}{2}-2}\!\left[1+
\!\frac{\Delta(x)}{2}\!+\! \frac{\Delta'(x)}{2} x\log x \right]\!dx =\!\frac{8\pi G}{3H_1^2}\rho.
\end{equation}
Further manipulation leads to
\begin{equation}
-\left[x^{\frac{\Delta(x)}{2}-1} + 2\!\int\! x^{\frac{\Delta(x)}{2}-2}\, dx \right]=\frac{8\pi G}{3H_1^2}\rho\,,
\label{ModF2bis}
\end{equation}
which represents the first Friedmann equation derived from the generalized Barrow entropy with a varying entropic exponent $\Delta(x)$. 
One immediately finds that, for $\Delta={\rm const}$, it takes a form similar to that obtained in Refs.~\cite{Saridakis:2020lrg,Sheykhi:2022jqq}.

\item As emphasized previously, if $\Delta$ encodes quantum effects in the context of gravity, a more physically consistent treatment 
can be achieved by considering a redshift-dependent parametrization and adopting the relation $a\equiv(1+z)^{-1}$; see, e.g., Refs.~\cite{Nojiri:2019skr,Basilakos:2023seo}. Since the Hubble rate plays the role of the cosmological variable, in lieu of the redshift, we propose
\begin{equation}
\label{eq:Delta}
    \Delta(x) = \Delta_0 + \frac{\Delta_1}{\ln x}\,,
\end{equation}
where $\Delta_0$ and $\Delta_1$ are suitable constants.

The physical motivation behind this choice is based on the fact that cosmic distances are inversely proportional to the Hubble scale. Hence, one can employ a simple Taylor expansion in inverse powers of the Hubble parameter, truncated at first order. However, this choice does not match quantum gravity results, which suggest a logarithmic correction to the Bekenstein--Hawking entropy~\cite{Callan:1970yg,Symanzik:1970rt,Wilson:1971bg,Reuter:1996cp,Dagotto:1989gp}. To reconcile these two aspects, the logarithmic form of Eq.~\eqref{eq:Delta} provides a natural representation of both, i.e., we adopt a first-order truncated expansion, with the logarithmic variable depending on $H$ through $x$.

\item A third motivation that reinforces the previous two points is the fact that a similar dependence arises from quantum fluctuations and radiative corrections encoded in the renormalization-group flow. Indeed, in semiclassical gravity and quantum field theory in curved spacetime, it emerges from one-loop effects and trace anomalies, while in quantum-gravitational frameworks, such as asymptotically safe gravity, it reflects the scale dependence of the effective gravitational couplings~\cite{Solodukhin:2011gn}.
\end{enumerate}

In view of the above considerations, we emphasize that Eq.~\eqref{eq:Delta} is singular at $\ln x=0$, corresponding to $x=1$ and hence to $H=H_1$. Therefore, the proposed running exponent should be interpreted as an effective sub-Planckian description, valid as long as $H\ll H_1$, or equivalently $x\gg 1$.
Importantly, this condition is amply satisfied throughout the entire observational range considered in the present analysis. Indeed, for the best-fit value $H_0\simeq69.27\,{\rm km\,s^{-1}\,Mpc^{-1}}$, one finds $H_0/H_1\sim10^{-61}$ and $x_0\sim10^{122}$. The hierarchy remains extremely large also at the photon-decoupling and neutrino-transition epochs. Assuming radiation domination and the best-fit model parameters, the formal condition $H=H_1$ would only be approached at redshifts of order $z\sim10^{31}$, far beyond the domain probed by the datasets used here and close to the regime in which a semiclassical cosmological description is no longer expected to be reliable\footnote{According to these considerations, no singularity is encountered in the numerical analysis, while any extrapolation of Eq.~\eqref{eq:Delta} toward Planckian or trans-Planckian epochs lies outside the domain of validity of the present effective framework.}.

Further, compared to the broader class of solutions proposed in Ref.~\cite{Nojiri:2019skr}, Eq.~\eqref{eq:Delta} constitutes a minimal running extension of the constant-exponent scenario. In particular, it introduces only one additional parameter beyond the standard Barrow model while preserving the correct constant-exponent limit for $\Delta_1\to0$.

\subsection{Assessing cosmic observations through the extended Barrow entropy}

In order to explain the entire cosmic expansion history, our novel parametrization has to be directly compared with previous phenomenological dark energy scenarios~\cite{Luongo:2024fww,PhysRevD.111.023512}. For example, unlike commonly adopted phenomenological parameterizations, such as the widely used $w_0w_a$CDM model~\cite{2001IJMPD..10..213C,Linder:2002et,Carloni:2025dqt}, our paradigm is not restricted to a limited redshift range. Hence, our choice ensures a smooth interpolation across cosmic epochs in a unified framework~\cite{Basilakos:2023seo}, allowing for a direct comparison with different data sets.

\begin{figure}
\centering
\includegraphics[width=0.98\hsize,clip]{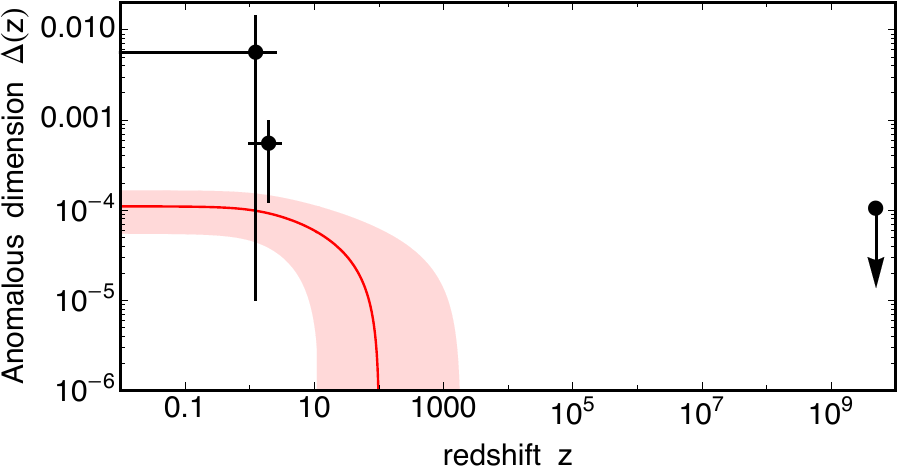}
\vspace{-0.15cm}
\caption{The $\Delta(z)$ parametrization (red curve with $1\sigma$ error bands), compared to the observational constraints at various epochs (black data with error bars and upper limits). The transformation from $x$ to $z$ is performed using Eq.~\eqref{eq:genx}. The selected $\Delta_0$ and $\Delta_1$ are the best fit values listed in Table~\ref{tab:best}.}
\label{fig1}
\end{figure}

To assess the observational viability of the proposed framework, we compare Eq.~\eqref{eq:Delta} with the currently available estimates of $\Delta$ at different cosmic epochs. The qualitative behavior is illustrated in Fig.~\ref{fig1}. Late-time estimates are comparatively better constrained~\cite{Leon:2021wyx,Yarahmadi:2024oqv,Luciano:2025elo,Luciano:2025hjn}, whereas early-time constraints are currently limited to upper bounds~\cite{Barrow:2020kug,Luciano:2022pzg,Luciano:2023roh}. Within these presently available constraints, our parametrization remains consistent with the data.

At this stage, plugging Eq.~\eqref{eq:Delta} into Eq.~\eqref{ModF2bis} determines the Hubble parameter,
\begin{equation}
\label{eq:genH}
    \frac{H(z)}{H_0} = \left[\left(\frac{2-\Delta_0}{2+\Delta_0}\right) \left(\frac{H_0}{H_1}\right)^{\Delta_0} e^{-\frac{\Delta_1}{2}} \Omega(z) \right]^{\frac{1}{2-\Delta_0}}\,,
\end{equation}
and the expression for $x$, required to plot $\Delta$ as a function of redshift in Fig.~\ref{fig1}, is given by
\begin{equation}
\label{eq:genx}
    x = \left[\left(\frac{2-\Delta_0}{2+\Delta_0}\right) \left(\frac{H_0}{H_1}\right)^2 e^{-\frac{\Delta_1}{2}} \Omega(z) \right]^{\frac{2}{\Delta_0-2}}\,.
\end{equation}
In Eqs.~\eqref{eq:genH}--\eqref{eq:genx}, the total cosmological density parameter $\Omega(z) = {8\pi G}\rho(z)/(3 H_0^2)$ can be decomposed into the contributions of the individual fluid components driving the cosmic dynamics, such that $\Omega(z) = \sum_i \Omega_i(z)$. Specifically, assuming a spatially flat universe, the total density parameter can be written as
\begin{equation}
\Omega(z) = \Omega_m(1+z)^3 + \Omega_\gamma(1+z)^4 + \Omega_\nu f(z) + \Omega_\Lambda\,,
\end{equation}
where $i=\{m,\gamma,\nu,\Lambda\}$ indicate pressureless matter, photons, neutrinos and cosmological constant, respectively. 

In particular, the energy density of neutrinos reads
\begin{equation}
\label{neutrinos}
  f_\nu(z) =
  \frac{\Omega_\gamma}{\Omega_\nu}
  \left[\frac{T_\nu(z)}{T_0}\right]^4 \sum_i \frac{I(r_i)}{3}\,,
\end{equation}
where $\Omega_\gamma = 8\pi a G T_0^4/(3 H_0^2)$ is the present-day photon density parameter, $T_0=2.34\times10^{-4}$~eV is the current CMB temperature and 
$T_\nu(z)=T_0 (4/11)^{1/3}N_{\rm eff}^{1/4}(1+z)$ denotes an effective neutrino temperature. Here, $N_{\rm eff}=3.046$ is the effective number of relativistic neutrino species, while $r_i=m_i/T_\nu(z)$ depends on the mass $m_i$ of each neutrino species.
The numerical integral $I$ is given by
\begin{equation}
  I(r_i) = \frac{15}{\pi^4} \int_0^\infty \frac{\sqrt{x^2+r_i^2}}{e^x+1} x^2 dx\,.
\end{equation}
Massless neutrinos, for which $I(0)=7/8$, behave as radiation, with
$f_\nu(z)\propto(1+z)^4$. Massive neutrinos instead behave as
pressureless matter at late times, when $I(r)\propto r\propto(1+z)^{-1}$,
implying $f_\nu(z)\propto(1+z)^3$, while they can be treated as
effectively massless at sufficiently early times. We assume a total
neutrino mass $\sum_i m_i = 0.06\,\mathrm{eV}$, distributed among one
massive and two massless species, consistently with neutrino oscillation
constraints \cite{Gonzalez-Garcia:2021dve}.

Finally, by imposing the condition $ H(0) \equiv H_0 $, the density parameter of the cosmological constant can be defined from Eq.~\eqref{eq:genH} as
\begin{equation}
\label{eq:Lambda}
\Omega_\Lambda = \left(\frac{2+\Delta_0}{2-\Delta_0}\right) \left(\frac{H_1}{H_0}\right)^{\Delta_0} e^{\frac{\Delta_1}{2}}-\Omega_m-\Omega_\gamma-\Omega_\nu\,.
\end{equation}

\section{Numerical analysis}
\label{NumAn}
Here, we describe the data sets employed in the Monte Carlo--Markov chain (MCMC) analysis and subsequently define the corresponding cosmological observables relevant for parameter estimation. Each data set is presented together with the observational methods and assumptions underlying the measurements, as well as the statistical framework used to construct its likelihood function. Finally, we summarize the results obtained from the combined analysis of all the data sets.

\subsection{Cosmic data sets}

The data sets used in this analysis, together with the definitions of the corresponding observables, are described below.

\begin{itemize}
\item[-] {\bf OHD.}
These Hubble rate measurements $H_i$ (see Appendix~\ref{AppA}) are determined spectroscopically via the identity $H(z)=-(1+z)^{-1}\Delta z/\Delta t$ \citep{2002ApJ...573...37J}.
The differences in age, $\Delta t$, and redshift, $\Delta z$, are obtained from pairs of galaxies that are assumed to have formed at the same epoch and to have rapidly exhausted their gas reservoirs, thereby evolving passively. These measurements are affected by large systematic errors (typically $20$-$30\%$), primarily due to factors such as star formation history, stellar age estimates, formation timescales, chemical composition and related effects \citep{Moresco:2022phi}.

Assuming Gaussian distributed errors $\sigma_{H_i}$, the corresponding log-likelihood function is given by
\begin{equation}
\label{loglikeOHD}
  \hspace{2em}\ln \mathcal{L}_{\rm O} = -\frac{1}{2} \sum_{i=1}^{34}\left\{\!\frac{[H_i-H(z_i)]^2}{\sigma_{H_i}^2}\!+\!\ln(2\pi\sigma_{H_i}^2)\!\right\}\!,
  \end{equation}
where $H(z)$ is described by Eq.~\eqref{eq:genH}.
\item [-] {\bf Pantheon+ SNe~Ia.} This catalog (see Appendix~\ref{AppA} for information) comprises 18 different samples 
and incorporates the SH0ES Cepheid host distance anchors \citep{Brout:2022vxf}, yielding a total of $1701$ SNe Ia  with redshift coverage $z \in [0, 2.3]$ \citep{2022ApJ...938..113S}.
Labeling the inferred SN distance moduli with $\mu_i$, the corresponding model predictions are defined by
\begin{equation}
\label{eq:SNdl}
\mu_{\rm th}(z_i) = m_i - M = 5\log[D_{\rm L}(z_i)] +25\,,
\end{equation} 
where $m_i$ are the rest-frame $B$-band apparent magnitudes, $M$ is the rest-frame $B$-band absolute magnitude, and the luminosity distance is defined as 
\begin{equation}
\label{Ldistance}
D_{\rm L}(z) =
(1+z)\int_0^z\frac{dz^\prime}{H(z^\prime)}.
\end{equation}
Following Ref.~\cite{SNLS:2011lii}, the absolute magnitude $M$ is treated as a nuisance parameter and analytically marginalized to the value $M = \mathcal{B}/\mathcal{E}$, in order to maximize the SN~Ia log-likelihood. As a result, the latter becomes
\begin{equation}
\label{eqn:chimarg}
 \hspace{2em}\ln{\mathcal{L}_{\rm S}} = -\frac{1}{2}\left[\mathcal A + \ln\left(\frac{\mathcal E}{2 \pi}\right) - \frac{\mathcal B^2}{\mathcal E}\right],
\end{equation}
where $\mathcal{A} \equiv \Delta m_i\, C_{ij}^{-1} \Delta m_j$, $\mathcal{B} \equiv \Delta m_i\, C_{ij}^{-1} I_j$ and $\mathcal{E} \equiv I_i\, C_{ij}^{-1} I_j$. The covariance matrix $C_{ij}$ includes both statistical and systematic errors, $I_i$ is a unit vector and $\Delta m_i \equiv m_i - m_{\rm th}(z_i)$ denotes the residual between the observed and model-predicted apparent magnitudes, which are given by 
\begin{equation}
\label{eq:SNmagn}
 m_{\rm th}(z_i) = 5\log\left[D_{\rm L}(z_i)\right] + 25.
\end{equation}
\item[-]{\bf DR2 DESI-BAO.} The DR2 of DESI-BAO data set is shown in Appendix~\ref{AppA} 
and includes six transverse distances $D_{\rm M}$, six Hubble rate distances $D_{\rm H}$ and seven angle-averaged distances $D_{\rm V}$, all normalized by the comoving sound horizon $r_d = r_s(z_d)$ at the baryon drag redshift $z_d$ \citep{DESI:2025zgx}. The corresponding expressions are given by
\begin{subequations}\label{distances}
\begin{align}
\frac{D_{\rm M}(z)}{r_d} &= \frac{D_{\rm L}(z)}{r_d(1+z)},\\[1mm]
\frac{D_{\rm H}(z)}{r_d}& = \frac{1}{r_d H(z)},\\[1mm]
\frac{D_V(z)}{r_d} &= \frac{\left[z D_{\rm H}(z) D_{\rm M}^2(z) \right]^{1/3}}{r_d}.  
\end{align}
\end{subequations}
For modeling $r_{\mathrm{d}}$, we adopt the highly accurate non-parametric reconstruction proposed in Ref.~\cite{Aizpuru:2021vhd}, which expresses the comoving sound horizon as
\begin{equation}
\label{eq:neutrino}
 \hspace{2em} r_\mathrm{d} = \frac{b_1~e^{b_2\left(b_3+\omega_{\nu}\right)^2}}{b_4~\omega_b^{b_5}+b_6~\omega_m^{b_7}+b_8\left(\omega_b\hspace{1mm}\omega_m\right)^{b_9}}~ \mathrm{Mpc}\,,
\end{equation}
where $\omega_{\nu} = h^2 \Omega_\nu$ and $\omega_m = h^2 \Omega_m$ are the comoving density parameters for massive neutrino species and total matter, respectively, while the comoving baryon density is fixed to $\omega_b = 0.02237 \pm 0.00015$~\cite{Planck:2018vyg}. In all the expressions $h=H_0/(100~{\rm km/s/Mpc})$ is the dimensionless Hubble parameter.
The numerical coefficients $b_i$ are
\vspace{-1em}
\begin{center}
\begin{minipage}{0.92\linewidth}
\begin{equation}
\nonumber
\begin{array}{lll}
b_1 = 0.0034917, & b_2 = -19.972694, & b_3 = 0.000336186,\\[1mm]
b_4 = 0.0000305, & b_5 = 0.22752,    & b_6 = 0.0000314257,\\[1mm]
b_7 = 0.5453798, & b_8 = 374.14994,  & b_9 = 4.022356899.
\end{array}
\end{equation}
\end{minipage}
\end{center}
The expression for the comoving sound horizon in Eq.~\eqref{eq:neutrino} is calibrated assuming the standard pre-recombination expansion history. Its applicability within the present framework is discussed in Appendix~\ref{AppAbis}.

The total BAO log-likelihood is therefore given by
\begin{equation}
   \hspace{2em} \ln\mathcal{L}_{\rm B} = -\frac{1}{2} \sum^{N_Y}_{i=1}\left\{\frac{[Y_i-Y(z_i)]^2}{\sigma_{Y_i}^2}\!+\!\ln(2\pi\sigma_{Y_i}^2)\right\},
\end{equation}
where $Y(z_i) = \{ D_M(z_i)/r_d, D_H(z_i)/r_d, D_V(z_i)/r_d \}$ are defined in Eqs.~\eqref{distances}, $Y_i$ and $\sigma_{Y_i}$ denote the observed DR2 data and their errors, and $N_Y = \{6, 6, 7\}$ is the number of data points for each of the three distance measures, respectively\footnote{The use of DESI BAO data seems to indicate a mild dark energy evolution, albeit alternative views and reformulations of their use are often discussed in the literature, see e.g. Refs. \cite{Luongo:2024fww,Carloni:2024zpl,Alfano:2024jqn,Alfano:2025gie,Alfano:2024fzv,Alfano:2025fyq,Colgain:2025fct,Colgain:2025nzf,Berti:2025phi,GuptaChoudhury:2026gsl,GuptaChoudhury:2026hwk}.}.
\item[-]{\bf CMB shift parameters.}
These parameters
\begin{subequations}
\label{eq:CMB}
    \begin{align}
\label{eq:R}
\mathcal R(z_{*}) &= D_{\rm M}(z_\star) H_0 \sqrt{\Omega_m},\\[1mm]
\label{eq:lA}
l_{\rm A}(z_{*}) &= \pi \frac{D_{\rm M}(z_\star)}{r_\star},
\end{align}
\end{subequations}
are related to the acoustic angular scale at the recombination redshift $z_\star= 1089.92 \pm 0.25$ \citep{Planck:2018vyg}, i.e., $\theta_\star = r_\star/D_{\rm M}(z_\star)$, where $r_\star=r_s(z_\star)$ is the comoving sound horizon $r_s$ at that epoch, given by
\begin{equation}
\label{eq:0}
r_\star = \frac{1}{\sqrt{3}} \int_{z_{\star}}^{\infty} \frac{dz}{H(z)\sqrt{1 + R(z)}}.
\end{equation}
The parameter $R(z)=(3/4)(\omega_b/\omega_\gamma)(1 + z)^{-1}$ depends on the comoving density parameters of baryons and photons, where $\omega_\gamma = h^2 \Omega_\gamma$.

Using the measured $\mathcal{R} = 1.7502 \pm 0.0046$ and $l_A = 301.471^{+0.089}_{-0.090}$~\citep{Chen:2018dbv}, the total CMB log-likelihood is
\begin{equation}
\hspace{2em}    \ln\mathcal{L}_{\rm C} = -\frac{1}{2} \sum_{X}\left\{\frac{[X-X(z_\star)]^2}{\sigma_{X}^2}\!+\!\ln(2\pi\sigma_X^2)\right\},
\end{equation}
where $X=\{\mathcal R,l_A\}$, $\sigma_{X}$ denotes the errors, and $X(z_\star)= \{\mathcal R(z_\star), 
l_A(z_\star)\}$ are given in Eqs.~\eqref{eq:CMB}.
\end{itemize}

Based on all the data sets described above, the model parameters $\{H_0, \Omega_m, \Delta_0, \Delta_1\}$ are determined by maximizing the total log-likelihood function, defined as
\begin{equation}
\label{eq:totlike}
\ln\mathcal{L} = \ln \mathcal L_{\rm O}  + \ln \mathcal L_{\rm S} + \ln \mathcal L_{\rm B} + \ln \mathcal L_{\rm C}.
\end{equation}

Before proceeding with the parameter estimation, it is worth noting two aspects that distinguish the present analysis from previous studies based on scale-dependent power-law extensions of the Bekenstein entropy. 
From the theoretical perspective, we adopt a minimal scale-dependent parametrization of the Barrow exponent, described by $\Delta_0$ and $\Delta_1$, which yields an analytic expression for the Hubble rate and continuously recovers the $\Lambda$CDM model in the limit $\Delta_0=\Delta_1=0$. 

From the observational perspective, our analysis is extended beyond SNe~Ia and direct $H(z)$ measurements by combining OHD, Pantheon+ SNe~Ia with SH0ES Cepheid-host anchors, DESI DR2 BAO data, and CMB shift parameters. This allows us to test the model over a wider range of cosmic epochs and to constrain jointly the parameters $\{H_0,\Omega_m,\Delta_0,\Delta_1\}$, while directly comparing its statistical performance with that of $\Lambda$CDM.

\section{Physical results}\label{sezione5}

The results are summarized in Table~\ref{tab:best}, where we compare the proposed entropy parametrization with the $\Lambda$CDM model. The corresponding posterior distributions are shown in the contour plots of Fig.~\ref{fig:cont}.

Given the large total number of data points $N=1756$, the model free parameters $p$, and the maximum value $\mathcal L_0$ of Eq.~\eqref{eq:totlike}, we can assess the best fit model by using the Bayesian Information Criterion (BIC) \cite{2007MNRAS.377L..74L}
\begin{equation}
{\rm BIC}=-2\ln \mathcal L_0+p\ln N\,.
\end{equation}
Our analysis reveals that the $\Lambda$CDM paradigm is the best fitting (fiducial) model, yielding the lowest BIC. In contrast, the entropy-based model exhibits a larger BIC, with its difference relative to the $\Lambda$CDM model being $\Delta{\rm BIC} > 6$. According to the conventional interpretation, this result provides statistical evidence in favor of the fiducial model. This finding is in agreement with the conclusions reported in Ref.~\cite{Luciano:2025hjn} for the case of the standard (i.e., constant $\Delta$) Barrow entropy-based cosmology, constrained using combined SN, OHD and BAO datasets.

Despite the statistical preference for the simpler $\Lambda$CDM model, the proposed framework remains fully compatible with the current observations and represents a consistent extension of the standard cosmological scenario. It is therefore worthwhile to further investigate its phenomenological implications, particularly its potential to address some of the open issues of the concordance model. As a first step, using the results reported in Table~\ref{tab:best}, we compute the anomalous dimension at $z=0$, obtaining $\Delta(0)=1.09^{+0.55}_{-0.55}\times10^{-4}$. This result is consistent with recent low-redshift estimates presented in Refs.~\cite{Leon:2021wyx,Yarahmadi:2024oqv}.
\begin{table}
\centering
\setlength{\tabcolsep}{.65em}
\renewcommand{\arraystretch}{1.3}
\begin{tabular}{lll}
\hline\hline
Parameter               & Barrow Cosmology
                        & $\Lambda$CDM \\
\hline\hline
$H_0~({\rm km/s/Mpc})$  &        
$69.27^{+0.62(0.99)}_{-0.65(1.00)}$  & $68.31^{+0.28(0.45)}_{-0.26(0.43)}$  \\[0.5mm]
$\Omega_m$              & $0.310^{+0.009(0.016)}_{-0.010(0.015)}$ & $0.302^{+0.003(0.006)}_{-0.004(0.007)}$ \\[0.5mm]
$\Delta_0\ (\times10^3)$ & 
$2.40^{+1.30(2.30)}_{-1.30(2.10)}$ 
                        & $0$ \\[0.5mm]
$\Delta_1$              & 
$-0.65^{+0.35(0.58)}_{-0.38(0.65)}$    
                        & $0$ \\[0.5mm]
$\ln\mathcal L_0$       & $-1016.03$
                        & $-1020.02$\\[0.5mm]
BIC                     & $2061.92$                            & $2054.98$\\[0.5mm]
$\Delta$BIC             & $6.94$                               & $0$ \\[0.5mm]
\hline
\end{tabular}
\caption{Best fit parameters with $1\sigma$ ($2\sigma$) errors and maximum log-likelihood values for the Barrow cosmology and the $\Lambda$CDM model. The last two rows report the BIC statistical criterion values and the corresponding differences.}
\label{tab:best}
\end{table}

\begin{figure}
\centering
\includegraphics[width=0.98\hsize,clip]{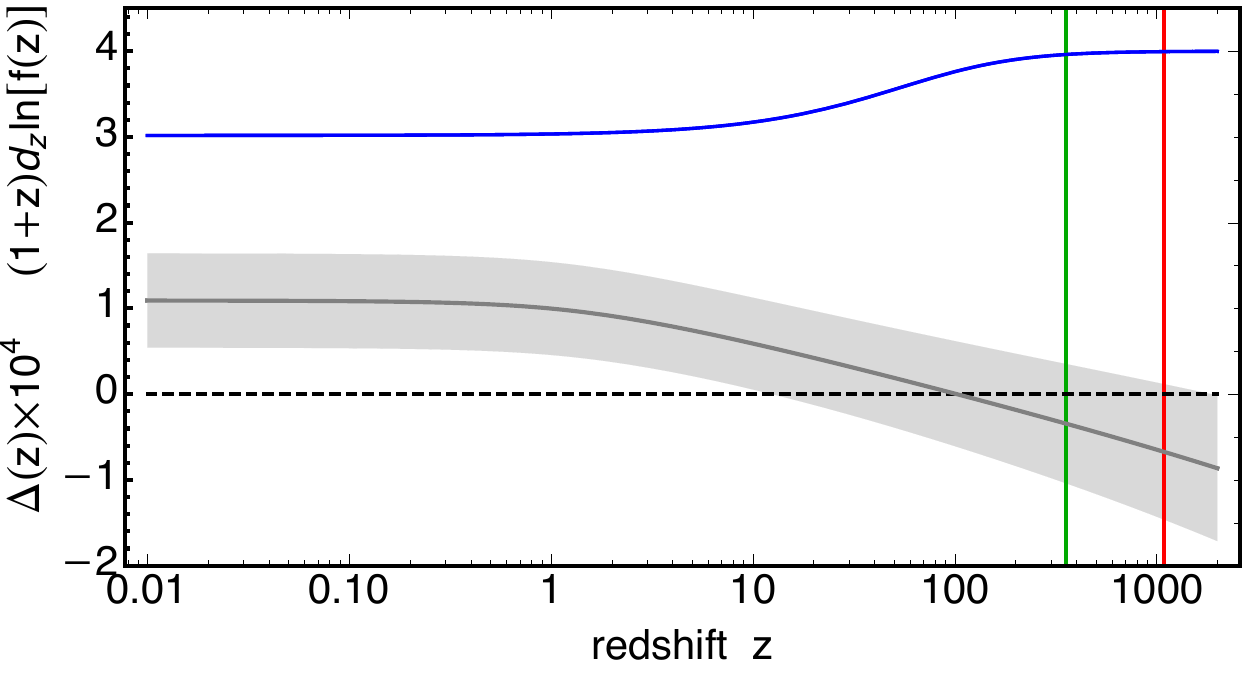}
\vspace{-0.15cm}
\caption{Behavior of $\Delta(z)$ (magnified by a factor of $10^4$) as inferred from the results in Table~\ref{tab:best} (gray curve with error bands), along with the reference value $\Delta = 0$ (dashed horizontal line). For comparison, the matter-to-radiation transition function for neutrinos, $(1+z)\, d_z \ln[f(z)]$, is shown (blue curve, upper part of the plot). Vertical red and green lines indicate the redshifts of photon decoupling ($z_\star$) and neutrino transition to the relativistic regime ($z_\nu$), respectively.
}
\label{fig2}
\end{figure}

Next, Fig.~\ref{fig2} shows that the parametrization introduced in Eq.~\eqref{eq:Delta} naturally interpolates between two regimes,
\begin{itemize}
    \item[-] $\Delta(z)>0$ for $z<z_{\rm tr}$,
    \item[-] $\Delta(z)<0$ for $z>z_{\rm tr}$,
\end{itemize}
through a smooth transition occurring at a redshift
$z_{\rm tr}=102^{+1770}_{-90}$. Within the inferred uncertainties, this transition is compatible both with the photon decoupling redshift $z_\star$ and with the characteristic redshift $z_\nu\simeq356$, corresponding to the epoch at which neutrinos become non-relativistic, defined by the condition $T_\nu(z_\nu)\approx\sum_i m_i$.

Focusing on the second scenario, one can define a matter-to-radiation transition function for neutrinos that describes the evolution from a density $\propto (1+z)^3$ to one scaling as $\propto (1+z)^4$. Specifically, we introduce
\begin{equation}
    \frac{d\ln[\Omega_\nu f(z)]}{d\ln(1+z)} = (1+z)\, d_z \ln[f(z)]\,,
\end{equation}
where $d_z \equiv d/dz$. By construction, this function takes the value $3$ for matter density and $4$ for radiation density. As shown in Fig.~\ref{fig2}, for neutrinos it assumes values in the range $[3,4]$ within the redshift interval $z \in [10,500]$, which is consistent with the estimated $z_{\rm tr}$.

Last but not least, we have explicitly verified that the theoretically admissible range,
$-1<\Delta(z)\leq1$, is satisfied throughout the entire redshift interval
covered by the observational datasets. For the best-fit values reported in
Table~\ref{tab:best}, the running exponent remains of order $10^{-4}$,
varying from $\Delta(0)\simeq10^{-4}$ at the present epoch to
$\Delta(z_\star)\simeq-10^{-4}$ at photon decoupling.

Hence, the evolution inferred from the observational analysis remains well
within the theoretically allowed interval and corresponds to only a small
deviation from the Bekenstein--Hawking area law. We have also verified that
this conclusion is robust under variations of the model parameters within
their reported confidence intervals.

\subsection{The sign reversal of $\Delta$}

The behavior of the anomalous dimension $\Delta(z)$ encodes nontrivial information about the properties of the horizon and its underlying thermodynamic description. In particular, its sign reversal can be interpreted as the microscopic manifestation of a transition in the effective spacetime microstructure, reflecting a profound reorganization of its statistical and geometric degrees of freedom. According to Eq.~\eqref{Barrow}, $\Delta > 0$ corresponds to a super-area scaling of the entropy, signaling an enhancement of the effective number of gravitational degrees of freedom relative to the semiclassical case ($\Delta = 0$). Conversely, $\Delta < 0$ leads to a sub-area scaling, indicating a reduction in the number of accessible microstates.

It is therefore interesting to speculate on the possible physical origin of this behavior within our model. From a quantum gravity perspective, the early Universe at high redshift ($z \gg z_{\rm tr}$) is effectively characterized by a UV-dominated phase, in which higher curvature effects and high-energy phenomena govern the microscopic structure of spacetime. In this regime, a negative value of $\Delta$ may reflect the action of restrictive quantum gravity mechanisms -- such as dimensional reduction~\cite{tHooft:1993dmi,Carlip:2000nv,Horava:2009if} and topological constraints~\cite{Witten:1988ze,Hawking:1971tu} -- that suppress the number of admissible microstates.

Furthermore, on the particle physics side, this phase corresponds to an early-Universe 
epoch near the radiation-dominated era, characterized by a hot relativistic plasma 
composed of photons, baryons and neutrinos in thermal equilibrium. The dominance 
of relativistic degrees of freedom, together with diffusive and free-streaming 
processes, damps small-scale perturbations through mechanisms such as Silk damping 
and collisionless diffusion~\cite{Silk:1967kq,Bond:1980ha}, thereby constraining the 
available phase space of microscopic configurations. Given that quantum fields both 
backreact on spacetime geometry~\cite{Jacobson:1995ab,Padmanabhan:2010xh} and 
contribute to the microscopic origin of gravitational entropy through their 
entanglement entropy across the horizon~\cite{Bombelli:1986rw,Srednicki:1993im,
Solodukhin:2011gn}, the suppression of short-wavelength modes translates into a 
reduction of the effective horizon degrees of freedom.

Taken together, these considerations suggest that the emergence of a negative 
$\Delta$ at high redshift may be interpreted as the combined manifestation of 
quantum gravity effects and high-energy particle physics processes, providing a consistent 
phenomenological imprint of UV physics on the entropy-area relation.

On the other hand, as the Universe evolves toward lower redshift ($z \lesssim z_{\rm tr}$), 
the sign reversal to a positive $\Delta$ may signal the onset of a new phase in which 
long-wavelength gravitational modes and the inhomogeneities associated with late-time 
structure formation imprint multiscale distortions on the horizon surface. From the 
particle physics viewpoint, this epoch approximately corresponds to the 
post-recombination ``dark ages'', when photons have decoupled and gravitational 
clustering begins to generate correlations on progressively larger scales. The 
resulting horizon geometry may thus be envisioned as resembling a sphereflake-like 
fractal morphology, characterized by an effectively higher dimensionality and an 
expanded configurational space enriched by large-scale correlations.

Therefore, within our model, the transition across $z_{\rm tr}$ can be interpreted 
as marking the passage from an early-Universe phase characterized by 
microstate-suppressing behavior of the horizon degrees of freedom to a late-time 
regime of microstate-amplifying dynamics. This interpretation provides a coherent 
physical framework linking the scale-dependent entropic exponent to the evolution 
of the Universe from its primordial high-energy conditions to its large-scale 
structured state at late times.

\section{The Hubble tension}
\label{HubTen}

Here we investigate the possibility that our extended Barrow framework may alleviate the Hubble tension.\footnote{The usual strategies to address the Hubble tension involve introducing either an early dark energy field \cite{Kamionkowski:2022pkx} or additional barotropic fluids \cite{Carloni:2025jlk,Carloni:2026yut}. In this context, modifying the cosmic expansion history may induce regions of repulsive gravity \cite{Luongo:2014qoa}, thus providing an alternative approach.}
To this end, we note that the best-fit value of the Hubble constant reported in Table~\ref{tab:best}, $H_0 = 69.27^{+0.62}_{-0.65}$~km/s/Mpc, lies between the value inferred by the Planck Collaboration from CMB observations, $H_0^{\rm P} = 67.36^{+0.54}_{-0.54}$~km/s/Mpc \cite{Planck:2018vyg}, and that measured by the SH0ES Collaboration using the Cepheid-calibrated cosmic distance ladder, $H_0^{\rm R} = 73.04^{+1.04}_{-1.04}$~km/s/Mpc~\citep{2022ApJ...934L...7R}.

These two measurements lie at the core of the current Hubble tension in modern cosmology. It is therefore of particular interest to investigate whether this discrepancy can be alleviated within the framework of Barrow entropy.

\subsection{Early time constraints}

We first focus on the early-time determination of the Hubble constant.
In the standard approach, the acoustic scale $\theta_\star$ is measured and, assuming a $\Lambda$CDM background cosmology, the expansion history is extrapolated from the photon decoupling epoch to the present time, thereby inferring the value $H_0^{\rm P}$.

Here, we adopt an analogous strategy. Specifically, we equate the CMB shift parameters given in Eqs.~\eqref{eq:CMB}, computed using the best fit values from Barrow’s entropy model in Table~\ref{tab:best} (denoted by ``B''), with the corresponding shift parameters evaluated within the $\Lambda$CDM framework (denoted by ``$\Lambda$''):
\begin{subequations}
\begin{align}
\mathcal R^{\rm B}(H_0,\Omega_m,\Delta_0,\Delta_1,z_\star) &\equiv \mathcal R^\Lambda(H_0^\Lambda,\Omega_m,z_\star),\\[1mm]
l_A^{\rm B}(H_0,\Omega_m,\Delta_0,\Delta_1,z_\star) &\equiv l_A^\Lambda(H_0^\Lambda,\Omega_m,z_\star),
\end{align}
\end{subequations}
where, in $\mathcal{R}^\Lambda$ and $l_A^\Lambda$, we fix $\Omega_m$ to the best fit value reported in Table~\ref{tab:best}.  

Solving these relations for $H_0^\Lambda$ yields two constraints:
\begin{subequations}
\begin{align}
H_0^\Lambda&>62.70\ {\rm km/s/Mpc},\\[1mm]
\label{H0good}
H_0^\Lambda&=67.84^{+0.66}_{-0.66}\ {\rm km/s/Mpc}.
\end{align}
\end{subequations}
Remarkably, the constraint from  Eq.~\eqref{H0good} is in perfect agreement with $H_0^{\rm P}$.
Phrasing it differently, \emph{Barrow's entropy model purports $H_0=69.27$~km/s/Mpc, whereas $H_0^{\rm P}$ only results because of a low-redshift extrapolation got by imposing the $\Lambda$CDM background cosmology}.

\subsection{Late time constraints}

We now turn to the late-time determination of the Hubble constant.  
In this approach, $H_0$ is inferred directly from low-redshift observations, without assuming the $\Lambda$CDM background evolution to extrapolate from the early Universe.  We limit our analysis to SNe Ia of the Pantheon+ catalog with $z\leq0.01$. Their distance moduli are linked to the luminosity distance via Eq.~\eqref{eq:SNdl}. 
At such low redshift, the cosmographic series is employed to derive $H_0$ from the cosmic distance ladder as \cite{Dunsby:2015ers,Gruber:2013wua,Aviles:2012ay,Luongo:2012dv,Luongo:2013rba,Capozziello:2021xjw,Aviles:2016wel,Luongo:2011zz}
\begin{equation}
\label{eq:cosmo}
D_{\rm L}(z)\approx \frac{z}{H_0}\!\left(\!1 + \!\frac{1-q_0}{2}z + 
\!\frac{3q_0^2 + q_0 - j_0 - 1}{6} z^2\!\right),
\end{equation}
where $q_0$ and $j_0$ denote the deceleration and jerk parameters, respectively, evaluated at $z = 0$. In our model, these parameters take the following specific forms:
\begin{subequations}
\begin{align}
q_0&=-1 + \frac{3\Omega_m}{2+\Delta_0} \left(\frac{H_0}{H_1}\right)^{\Delta_0} e^{-\frac{\Delta_1}{2}},\\[1mm]
j_0&=1 + \frac{9\Omega_m^2\Delta_0}{(2+\Delta_0)^2} \left(\frac{H_0}{H_1}\right)^{2\Delta_0} e^{-\Delta_1}.
\end{align}
\end{subequations}

Following the procedure outlined in Eqs.~\eqref{eq:SNdl}-\eqref{eq:SNmagn}, in which the absolute magnitude $M$ is treated like a nuisance parameter, we extract a local bound on $H_0$ by fitting the luminosity distances of SNe Ia at $z\leq0.01$ using the expansion in Eq.~\eqref{eq:cosmo} and the best fit parameters $\{\Omega_m,\Delta_0,\Delta_1\}$ from Table~\ref{tab:best}. We find 
\begin{equation}
H_0^{\rm M}= 70.13^{+1.10}_{-1.10}\ {\rm km/s/Mpc},
\end{equation}
in agreement with the result of Table~\ref{tab:best}, but still in tension at $2.9\sigma$ confidence level with the local estimate $H_0^{\rm R}$.

Conversely, if we use the Cepheid-calibrated distance moduli $\mu^{\rm C}_i$ from SH0ES (see Appendix~\ref{AppA} for information), namely fixing the absolute magnitude $M$ based on the cosmic ladder, we obtain the corresponding calibrated luminosity distances from
\begin{equation}
\mu^{\rm C}_i=5\log [D_{\rm L}^{\rm C}(z_i)]+25.
\end{equation} 
Repeating the former methology, we fit the calibrated luminosity distances of SNe Ia at $z\leq0.01$ with Eq.~\eqref{eq:cosmo} and utilizing the values of $\{\Omega_m,\Delta_0,\Delta_1\}$ from Table~\ref{tab:best} we obtain a calibrated value of the Hubble constant 
\begin{equation}
H_0^{\rm C}= 74.36^{+1.15}_{-1.15}\ {\rm km/s/Mpc},
\end{equation}
which is in perfect agreement with $H_0^{\rm R}$.

The main conclusion from the above calculations can be summarized as follows: \emph{Barrow's entropy model purports $H_0=69.27$~km/s/Mpc when the absolute magnitude of SNe Ia is treated as a nuisance parameter, whereas $H_0^{\rm R}$ is a direct consequence of fixing $M$ based on the Cepheid-calibrated cosmic ladder}. 
This result, somehow, questions the validity of SNe Ia calibration or, rather, points to a still unknown flaw in extending the cosmic ladder at increasing values of the redshift.\footnote{See also Refs.~\cite{Carloni:2025jlk,Muccino:2026gvt}, for alternative views.}

\section{Final remarks}
\label{Concl}

In this work, we explored a modified cosmological framework arising from a generalized version of the Barrow entropy, in which the anomalous dimension varies and may naturally depend on the characteristic scale of the system, leading to a running behavior across different cosmological scales.
In particular, to motivate our choice, we proposed a first-order Taylor expansion in an inverse logarithmic variable, inspired by three main physical reasons. The first is that the corresponding entropy modification, namely the anomalous dimension, is expected to approach a constant as the Hubble rate increases.
The second consideration is that the Barrow entropy may be naturally motivated by quantum gravity arguments, encoding possible fractal deformations of the horizon geometry. Hence, logarithmic corrections are expected. Furthermore, such a scale dependence is typical when the renormalization group governs the variation of physical parameters.

Hence, combining the above motivations provides an alternative version of the Barrow scenario that can be applied to the first law of thermodynamics by employing the apparent horizon in an FRW Universe. As a consequence of this prescription, we modified the cosmic dynamics through the varying Barrow anomalous dimension, whose limiting cases naturally recover both the standard area law and the $\Lambda$CDM paradigm.

To do so, we introduced a specific ansatz for the Barrow running exponent that preserves observational consistency with late-time estimates, as well as with early-time upper bounds on $\Delta$ reported in the existing literature.
We then derived the modified expression for the Hubble rate and compared the proposed model with a combination of datasets, including OHD, Pantheon+ SNe Ia, DESI DR2 BAO, and CMB constraints. Although
the $\Lambda$CDM framework remains mildly favored, we found
that Barrow entropy exhibits a nontrivial phenomenology that may warrant further investigation.
Specifically, with our parametrization, we obtained
$\Delta(0) = 1.09^{+0.55}_{-0.55} \times 10^{-4}$, in agreement with recent low-redshift estimates reported in Refs.~\cite{Leon:2021wyx,Yarahmadi:2024oqv}.

Remarkably, $\Delta(z)$ exhibits a sign change from negative to positive values as $z$ decreases, with the transition occurring at
$z_{\rm tr} = 102^{+1770}_{-90}$, which is consistent with both the photon decoupling redshift and the redshift corresponding to the epoch when neutrinos become non-relativistic\footnote{Clearly, this transition redshift has nothing to do with the onset of cosmic acceleration; see, e.g., Ref.~\cite{Muccino:2022rnd}.}.
Moreover, this sign change is also consistent with previous studies suggesting that the positivity of the Barrow anomalous dimension, originally motivated in the context of black hole horizons, need not be preserved when the framework is extended to cosmology.

Interestingly, we addressed the possibility of alleviating the Hubble tension.
Indeed, including a thermodynamic source in the Friedmann equations modifies the Hubble expansion at different redshifts, thereby affecting its late-time behavior and, \emph{de facto}, the inferred value of the Hubble constant.
In this respect, we found the value $H_0 = 69.27^{+0.62}_{-0.65}~\mathrm{km\,s^{-1}\,Mpc^{-1}}$, suggesting a partial mitigation of the discrepancy between the Riess and Planck determinations \cite{Planck:2018vyg,2022ApJ...934L...7R}..

Hence, our investigation provides an advanced proposal in which the inclusion of a novel evolving term in the Barrow entropy can reconcile the early and late Universe within a unified framework.
Clearly, our analysis requires further investigation, and it would be interesting to perform additional comparisons between our model and established formalisms of quantum cosmology. More generally, extended thermodynamic descriptions of the Universe may be compared with our framework, with the aim of reconstructing \emph{a posteriori} the most appropriate functional form of $\Delta(z)$ beyond our approximation, for example in the context of alternative interpretations of dark matter; see, e.g., \cite{Luongo:2025iqq}. Nevertheless, future work will incorporate additional and complementary cosmological probes, such as high-redshift cosmic chronometers, gravitational-wave standard sirens, redshift-space distortion data, and strong-lensing time delays.

\section*{Acknowledgements}

The research of GGL is supported by the postdoctoral funding program of the University of Lleida.
GGL  also acknowledges  the contribution of the LISA CosWG and of  COST Actions CA21136 ``Addressing observational tensions in cosmology with systematics and fundamental physics (CosmoVerse)'', CA21106 ``COSMIC WISPers 
in the Dark Universe: Theory, astrophysics and experiments'', and CA23130 ``Bridging high and low energies in search of quantum gravity 
(BridgeQG)''.



\bibliographystyle{apsrev4-1}
\bibliography{Bib}

\newpage
\begin{appendix}

\section{References for the catalogs}
\label{AppA}

We here report the datasets employed in our analysis.


The OHD catalog, containing redshifts and measurements of $H(z)$, together with the corresponding statistical and systematic (or combined) errors, is taken from Ref.~\cite{Luongo:2024zhc}. 

The Pantheon+ catalog, comprising redshifts, apparent magnitudes, Cepheid-calibrated distance moduli and both statistical and systematic errors for all $1701$ SNe~Ia, is publicly available at \url{https://github.com/PantheonPlusSH0ES/DataRelease}. 

For BAO, we adopt the DESI~DR2 catalog, which includes measurements from the Bright Galaxy Survey (BGS), luminous red galaxies (LRG), emission-line galaxies (ELG), quasars (QSO), Lyman-$\alpha$ forest quasars (Ly$\alpha$-QSO) and the combined LRG+ELG sample. It provides six transverse comoving distances $D_{\rm M}$, six Hubble distances $D_{\rm H}$ and seven angle-averaged distances $D_{\rm V}$, each normalized by the sound horizon $r_d$, as reported in Ref.~\cite{DESI:2025zgx} (see also Refs.~\cite{Li:2025dwz,Zhou:2025nkb,Gialamas:2025pwv,Petri:2025swg}). 

Finally, the observational values of the CMB shift parameters adopted in this work are taken from Ref.~\cite{Chen:2018dbv}.


\section{Impact on the scale $r_d$}
\label{AppAbis}

The comoving sound horizon at the baryon drag epoch is
\begin{equation}
\label{eq:rd_integral}
r_d =
\frac{1}{\sqrt{3}} \int_{z_d}^{\infty}
\frac{dz}{H(z)\sqrt{1 + R(z)}}\,.
\end{equation}
While the modified entropy-area relation does not directly affect the photon--baryon sound speed, since the matter, photon, and baryon sectors retain their standard evolution, changes in $r_d$ can only arise through the modified $H(z)$, as well as through a possible shift in $z_d$.

We evaluate the magnitude of this effect by expanding Eq.~\eqref{eq:genH} around $\Delta_0\approx0$ and retaining only the leading contributions as $z\rightarrow z_d$, namely $\Omega_m(1+z)^3$ and $\Omega_r(1+z)^4$, where $\Omega_r=\Omega_\gamma+\Omega_\nu$.
We obtain
\begin{align}
\nonumber
H(z)\approx&\left\{1+\frac{\Delta_0}{2} \left[\ln\left(\frac{H_0}{H_1}\right)+ \ln \sqrt{\Omega_h(z)}-1-\frac{\Delta_1}{4}\right]\right\}\times\\
\label{eq:Hcompact}
&e^{-\frac{\Delta_1}{4}} H_0 \sqrt{\Omega_h(z)}
\end{align}
where $\Omega_h(z)=\Omega_m(1+z)^3+\Omega_r(1+z)^4$.
Using the parameters reported in Table~\ref{tab:best} and considering $z_d\in[1050,1070]$, we obtain $r_d\in[149.9,151.7]$~Mpc, implying a deviation of $\lesssim3\%$ with respect to the $\Lambda$CDM paradigm.

This result is also confirmed by Fig.~\ref{fig1}, where the running anomalous dimension becomes negligible around the baryon-drag and photon-decoupling epochs, suggesting that its influence on the early-time background is limited.

\section{Contour plots}
\label{AppB}

\begin{figure*}
{\includegraphics[width=0.65\hsize,clip]{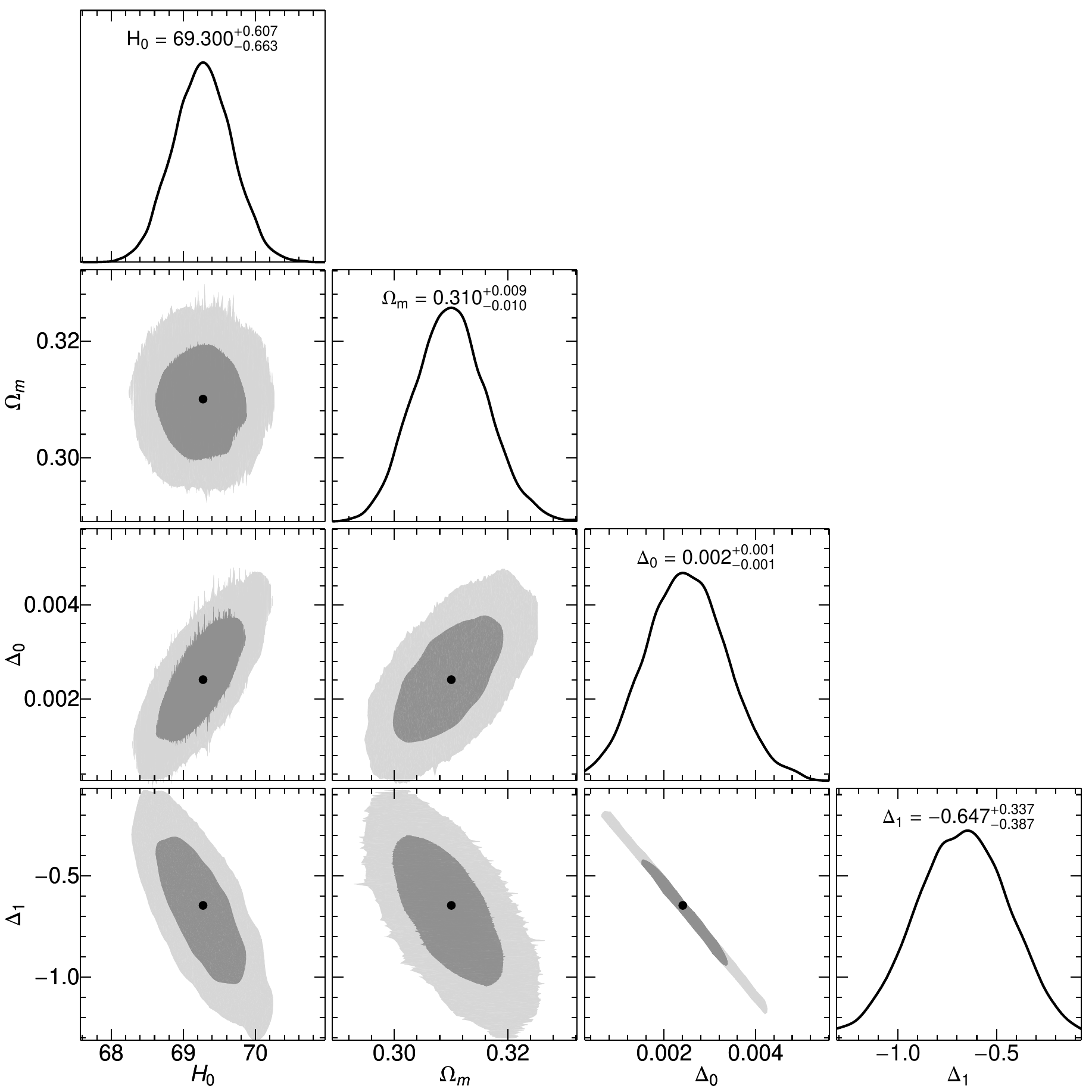} \hfill
\includegraphics[width=0.34\hsize,clip]{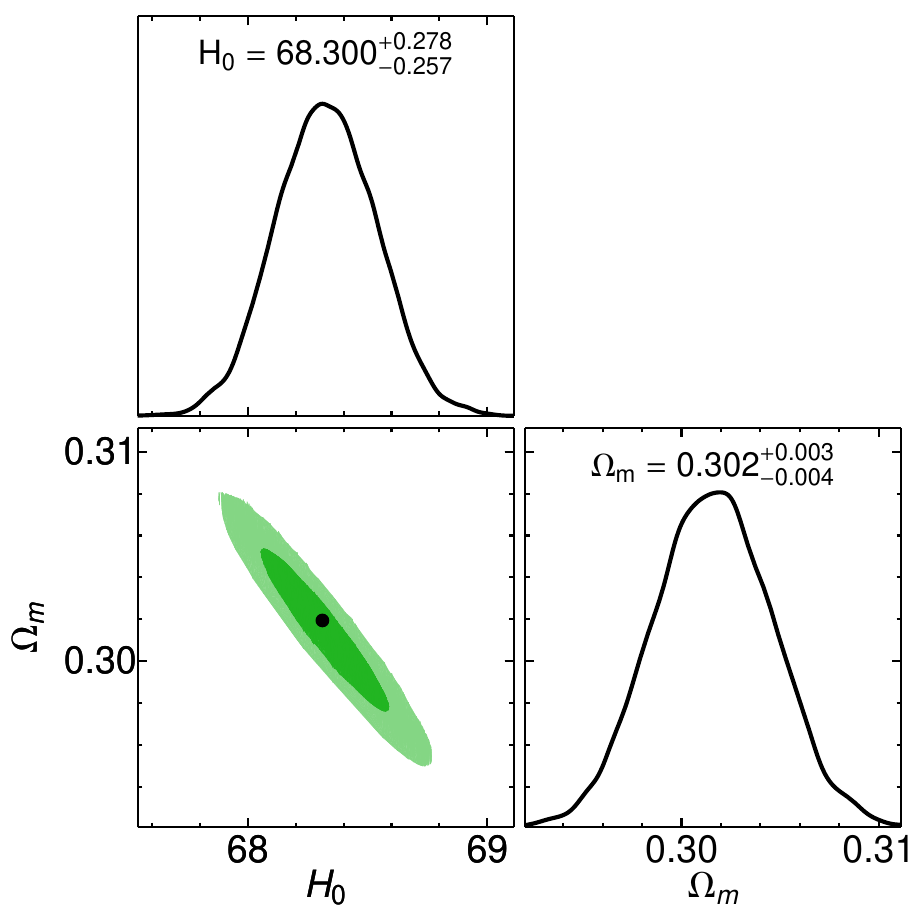}}
\caption{MCMC posteriors on the entropy model (left contours) and the $\Lambda$CDM paradigm (right contours). The dark (light) shaded regions correspond to $1\sigma$ ($2\sigma$) confidence levels, whereas the black dots correspond to the best fit values listed in Table~\ref{tab:best}.} 
\label{fig:cont}
\end{figure*}
The best-fit parameters obtained from our MCMC analysis are summarized in Table~\ref{tab:best}. The corresponding posterior distributions, together with the $1\sigma$ and $2\sigma$ confidence contours for both our entropy parametrization and the $\Lambda$CDM model, are shown in Fig.~\ref{fig:cont}..

\end{appendix}

\end{document}